\documentclass[prx,twocolumn,superscriptaddress,aps,amsmath]{revtex4-2}
\usepackage{amsfonts}
\usepackage{amssymb}
\usepackage{dsfont}
\usepackage{color}
\usepackage{graphicx}
\usepackage{xcolor}
\usepackage[normalem]{ulem}
\usepackage{braket}
\usepackage{makecell}
\graphicspath{{./imag/}}

\begin{document}

\newcommand{\av}[1]{\left\langle#1\right\rangle}
\newcommand{\id}{\mathds{1}}
\newcommand{\im}{\mathcal{J}}
\newcommand{\ds}{\mathbf{X}}
\newcommand{\llh}{\mathcal{L}}
\newcommand{\tr}{{\rm Tr}}
\newcommand{\n}{\hat{n}}
\newcommand{\env}{E}
\newcommand{\imp}{I}

\newcommand{\da}[1]{{\color{blue} #1}}
\newcommand{\il}[1]{{\color{purple} #1}}
\newcommand{\ms}[1]{{\color{teal} #1}}

\newcommand{\mpar}[1]{\marginpar{\small \it #1}}

\title{Scalable tomography of many-body quantum environments with low temporal entanglement
}
\author{Ilia A. Luchnikov}
\affiliation{Department of Theoretical Physics, University of Geneva, Quai Ernest-Ansermet 30, 1205 Geneva, Switzerland}
\affiliation{Quantum Research Center, Technology Innovation Institute, Abu Dhabi, UAE}
\author{Michael Sonnner}
\affiliation{Department of Theoretical Physics, University of Geneva, Quai Ernest-Ansermet 30, 1205 Geneva, Switzerland}
\author{Dmitry A. Abanin}
\affiliation{Department of Physics, Princeton University, Princeton, New Jersey 08544, USA}
\affiliation{Google Research, Brandschenkestrasse 150, 8002 Zurich, Switzerland}

\date{\today}

\begin{abstract}

Describing dynamics of a quantum system coupled to a complex many-body
environment  is a ubiquitous problem in quantum science. General
non-Markovian environments are characterized by their influence matrix~(IM)
-- a multi-time tensor arising from repeated interactions between the system
and environment. While complexity of the most generic IM grows exponentially
with the evolution time, recent works argued that for many instances of
physical many-body environments, the IM is significantly less complex. This
is thanks to area-law scaling of temporal entanglement, which quantifies the
correlations between the past and the future states of the system. However,
efficient classical algorithms for computing IM are only available for
non-interacting environments or certain interacting 1D environments. Here, we
study a learning algorithm for reconstructing IMs of large many-body
environments simulated on a quantum processor. This hybrid algorithm involves
experimentally collecting quantum measurement results of auxiliary qubits
which are repeatedly coupled to the many-body environment, followed by a
classical machine-learning construction of a matrix-product (MPS)
representation of the IM. Using the example of 1D spin-chain environments,
with a classically generated training dataset, we demonstrate that the
algorithm allows scalable reconstruction of IMs for long evolution times. The
reconstructed IM can be used to efficiently model quantum transport through
an impurity, including cases with multiple leads and time-dependent controls.
These results indicate the feasibility of characterizing long-time dynamics
of complex environments using a limited number of measurements, under the
assumption of a moderate temporal entanglement.

\end{abstract}
\maketitle
\section{Introduction}



Rapid experimental progress opened up a new era in which quantum processors are
increasingly capable of performing tasks that are challenging for classical
computers~\cite{Preskill2018NISQ, altman2021quantum, daley2022practical}. One
notable example of such a task, which received much attention, is the problem
of random circuit sampling~\cite{arute2019quantum, ZuchongzhiPRL21}. The advent
of powerful quantum processors calls both for further advances of classical
computational methods, and for identifying applications that are beyond the
reach of classical computers.

Non-equilibrium phenomena in quantum many-body systems~\cite{SchreiberMBLScience15, SchmiedmeyerScience2015GGE, KaufmanScience2016, AbaninRMP2019} are
a promising class of such applications, because dynamics often leads to rapid
growth of entanglement and computational complexity of system's wave function~\cite{daley2022practical}.
Traditional tensor-network methods~\cite{schollwock2011density,
cirac2021matrix}, efficient for ground states which typically feature low,
area-law entanglement, therefore generally struggle~\cite{vidal2004efficient,white2004real,paeckel2019time} to describe non-equilibrium phenomena such as evolution following quantum
quenches and quantum
transport, although notable advances were made for 1d systems~\cite{BertiniRMP2021}. 

\begin{figure*}
    \centering
    \includegraphics[width=\linewidth]{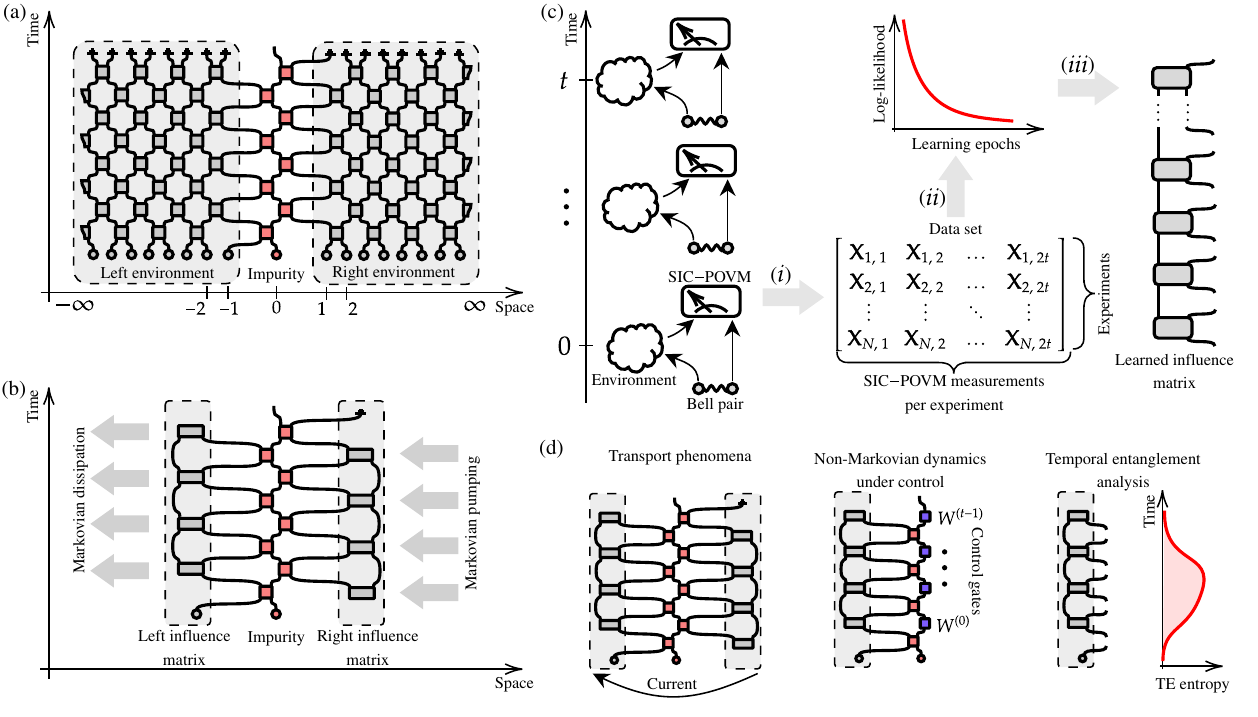}
    \caption{(a) A tensor network diagram representing discretized dynamics of an  impurity (red shaded area) connecting to two infinite one-dimensional reservoirs (environments). A particular example of reservoir considered below is an XX spin chain that maps onto free spinless fermions. Grey shaded parts of the diagram represent influence matrices of the environments. (b) A tensor network diagram approximating the same impurity dynamics, but effective compressed IMs. Dimensionality
    reduction is achieved by elimination of environment degrees of freedom which do not
    affect the impurity time evolution. (c) Key steps of the IM learning
    procedure: (\emph{i}) a data set is collected, by performing measurements on auxiliary degrees of freedom repeatedly interacting with an environment; (\emph{ii}) the data set is used by
    the learning algorithm that finds IM MPS ansatz matching the given data set
    the best; (\emph{iii}) A reconstructed MPS approximation of an IM is obtained, enabling
    numerical simulations of QIM dynamics. (d) A schematic illustration of example  applications of the learned IM, in particular: transport phenomena
    simulation, such as a current flow in QIIM; non-Markovian dynamics simulation under external control;
    temporal entanglement analysis.}
    \label{fig:intro_fig}
\end{figure*}

Recently, a new family of methods for quantum many-body dynamics, based on
compact matrix product state~(MPS) representations of the \emph{influence
matrix}~(IM) -- a generalized version of the Feynman-Vernon influence functional --
has emerged~\cite{banuls2009Matrix, lerose2021Influence, sonner2021Influence,
Chan21, frias-perez2022Light, lerose2023overcoming}. These methods are suitable
for computing dynamics of local physical observables, and, in contrast to more
conventional tensor-network techniques, their efficiency rests on moderate
\emph{temporal entanglement}~(TE) of the IM, which may be significantly lower
than spatial
entanglement~\cite{banuls2009Matrix,lerose2021Influence,lerose2021Scaling}.
These and related ideas, in particular, led to new efficient methods for
bosonic~\cite{strathearn2018efficient, cygorek2022simulation, fux2023tensor,
vilkoviskiy2024bound} and fermionic~\cite{thoenniss2023Efficient,
thoenniss2023Nonequilibrium, ng2023real} quantum impurity models (QIM)
out-of-equilibrium. An important class of QIM includes models of a small
quantum system coupled to a bath of (possibly infinitely) many non-interacting
degrees of freedom. Spin-boson QIMs are paradigmatic model for studying
non-Markovian dynamics. Fermionic QIMs have important applications in quantum
transport in mesoscopic systems~\cite{de2002out,pustilnik2004kondo}, and in dynamical mean-field theory methods for
modelling strongly correlated
materials~\cite{Georges96Dynamical,Kotliar06Eletronic}. A schematic of a QIM
arising in a transport setup with two environments is illustrated in
Fig.~\ref{fig:intro_fig}a.

Intuitively, moderate TE in QIM and some other non-equilibrium settings is due
to the fact that some of the information about the state of the impurity
propagates into the environment without affecting the future dynamics of the
impurity. This allows representing the environment with many degrees of freedom
in a much more compact form, by an evolution of a finite-dimensional
environment subject to Markovian dissipation, corresponding to irreversible
information loss (see Fig.~\ref{fig:intro_fig}b).

In another notable line of developments, a framework for describing
non-Markovian dynamics in terms of a \emph{process tensor}~(PT) --  a multi-time
generalization of a quantum channel -- has been
developed~\cite{pollock2018NonMarkovian}. Among other applications, the PT,
which is formally equivalent to the IM, was applied to define measures of
non-Markovianity~\cite{pollock2018operational}, and to characterize quantum
devices by performing PT tomography from experimental
data~\cite{white2020demonstration,White_2022}.

In this article, building on this recent progress, we propose to apply quantum
processors to the problem of constructing the IM of complex quantum many-body
environments. The schematic of the hybrid quantum-classical algorithm is
illustrated in Fig.~\ref{fig:intro_fig}c. Conceptually, its key steps are:
(\emph{i}) collecting measurement results of \emph{probe qubits} (or ancilla
qubits), which are repeatedly coupled to an evolving quantum many-body system,
the \emph{environment}; (\emph{ii}) feeding the collected data set to a
machine-learning algorithm which constructs an MPS representation for the IM
that matches the measurement results best according to the log--likelihood
function; (\emph{iii}) using the resulting IM to predict dynamics involving a
given environment, e.g. as a reservoir in a QIM.

We study the applicability of this hybrid approach for for reconstructing IM of
infinite quantum many-body environments at long evolution times. We note that
previous works~\cite{Milz_2018, white2020demonstration, White_2022}
investigated PT learning using a number of approaches, with some of
those approaches exhibiting an exponential overhead in the number of evolution  time steps, which precludes their scalability in the evolution time. However,
under the assumption of a finite Markov order of the PT which leads to a finite
bond dimension in the MPS ansatz for the IM, efficiency can be significantly
improved~\cite{white2023unifying, guo2022reconstructing}. Closest in spirit to the present work, Ref.~\cite{Guo_2020} explored
MPS-based tomography~\cite{Cramer_2010} of PT for interacting environments of
up to 6 interacting spins.

To assess the feasibility of learning the IM of quantum many-body
environments, we consider environments that are semi-infinite spin chains. The
training data for such environments are obtained using recently developed
efficient classical algorithms~\cite{sonner2021Influence,lerose2023overcoming},
yielding an accurate MPS representation of the IM for a large number of
time-evolution steps. As a key result, we demonstrate that the learning
procedure is scalable: in particular, we reconstruct IMs of several spin chains
in the thermodynamic limit for $60$ discrete time steps that correspond to an
IM with as many as $2^{240}$ entries. Reconstruction of such IMs requires
millions of samples, which are accessible with current quantum hardware such as
superconducting qubit processors~\cite{arute2019quantum,ZuchongzhiPRL21}.

While here we generate the dataset classically, this approach will be most useful for many-body environments where no classical algorithm to compute IM
exists, but temporal entanglement is sufficiently low, thus allowing for an approximation of an IM using MPS with a moderate bond dimension. Indeed,
recent works argued that the area-law temporal entanglement scaling is a
generic feature of systems with quasiparticles~\cite{lerose2023overcoming},
including interacting integrable systems~\cite{giudice2022Temporal}. However,
efficient algorithm for computing IM in such interacting systems are
only available for one-dimensional systems. In a large variety of many-body phases, low-energy excited states can be effectively described by quasiparticle excitations. Thus, we expect the
IM-learning to be applicable to such systems, in higher dimensions and varying
system geometry.


Once an MPS representation of an IM of a complex many-body environment has been
constructed, the dynamics of the QIM with that environment can be efficiently
analyzed classically, as illustrated in Fig.~\ref{fig:intro_fig}d. Crucially,
once an IM has been learned, QIM dynamics for arbitrary impurity parameters
(e.g. on-site interaction) and time-dependent controls become accessible. Below
we demonstrate this by reconstructing an IM of XX and XXX semi-infinite spin
chains, and using it to simulate several relevant examples of impurity
dynamics, including (\emph{i}) non-Markovian effects in dynamics of impurity
under external controls; (\emph{ii}) quantum transport through the impurity
connected to two reservoirs, whose IMs are learned separately. 

The paper is structured as follows: In Sec.~\ref{sec:im} we introduce the
influence matrix formalism as a way to compute local observables dynamics and
transport properties of many-body quantum systems. Further, we describe the
influence matrix learning method in Sec.~\ref{sec:influence_matrix_recovery}.
Next we discuss numerical results confirming the capabilities of the proposed
method in Sec.~\ref{sec:results}. Finally we discuss the results, challenges
and further development of the method in Sec.~\ref{sec:Outlook}.

\section{Influence matrix}
\label{sec:im}

\begin{figure}[ht]
	\centering
  \includegraphics[width=\linewidth]{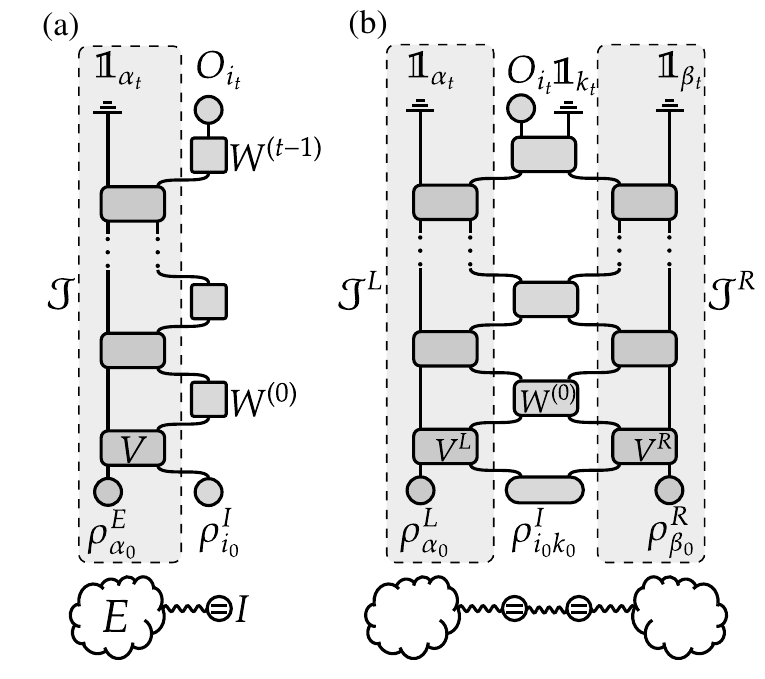}
  \caption{ (a) A tensor network diagram representing the dynamics of $\langle
  O(t)\rangle$, an observable of an impurity coupled with an
  environment. This diagram corresponds to the right-hand side of
  Eq.~\eqref{eq:obs_tn}, with a shaded area highlighting the
  IM of an environment $\im_{\{j_\tau,i_\tau\}_{\tau=0}^{t - 1}}$ (see Eq.~\eqref{eq:im}). Below the main tensor network
  diagram is a schematic representation of an impurity interacting with an
  environment. (b) A  tensor network diagram representing the dynamics of $\langle
  O(t)\rangle$ in a quantum transport setting. Here, $O$ is
   an observable of the left component of a two-qubit
  impurity coupled with two different environments. The superscripts $\im^L$
  and $\im^R$ refer to the left and right environments, respectively.}
  \label{fig:ims}
\end{figure}

We start by reviewing the influence matrix framework~\cite{lerose2021Influence}
(see also closely related PT formalism~\cite{pollock2018NonMarkovian}). The
IM is a multi-time tensor that contains complete information
regarding the effect of the environment on impurity dynamics. Formally, it can
be obtained by integrating out environment degrees of freedom. The complexity
of the IM can be significantly lower than the complexity of the combined
impurity-environment wave function, since not all the information regarding the internal state of environment is retained. This allows for compact
representations of IM in a number of non-equilibrium
settings~\cite{banuls2009Matrix,lerose2021Influence,lerose2021Scaling,sonner2021Influence,Chan21,frias-perez2022Light,lerose2023overcoming}. 

To define the IM, consider a QIM, where an impurity is a small quantum system
described by a finite-dimensional local Hilbert space, while an environment is
a many-body, possibly infinite-dimensional quantum system. We describe the
state of the impurity-environment system by a vectorized time-dependent density
matrix $\rho_{i\alpha}(t)$ (see Appendix~\ref{app:vectorization} for further
details on the vectorization). The first index refers to the impurity degrees
of freedom and the second index refers to the environment degrees of freedom.
For brevity of notation, we use Einstein's summation convention in what
follows. Our goal is to study the dynamics of the impurity by integrating out
the environment's degrees of freedom. Thus, we are interested in expectation
values of the following form:
\begin{align}
	\av{O(t)} = O_{i}\id_{\alpha}\rho_{i\alpha}(t),\label{eq:opdef}
\end{align}
where $O_i$ is a vectorized operator of a local observable $\hat{O}$ on the impurity
and $\id_\alpha$ is the vectorized identity operator that acts on the
environment degrees of freedom as the partial trace. The initial state of the
composite system $\rho_{i\alpha}(0)$ is taken to be a product state between the
initial state of the impurity $\rho^{\imp}_i$ and the environment $\rho^{\env}_\alpha$
\begin{align}
    \rho_{i\alpha} (0) = \rho^{\imp}_i \rho^{\env}_\alpha. 
\end{align}
We consider discrete time evolution of the composite system, which naturally
arises when the continuous time evolution of the QIM, defined by a
Hamiltonian or a Lindbladian, is discretized (trotterized). It is convenient to
decompose each time step of the evolution into two half-steps. The first
half-step, described by a quantum channel $V_{i\alpha,j\beta}$, represents an
evolution step due to environment's own dynamics, as well as due to the
system-environment interaction (see Fig.~\ref{fig:ims}). For simplicity, we
consider a time-translation invariant evolution of the environment, hence $V$
is constant (a generalization to the case when $V$ is time-dependent is
straightforward). During the second half-step of the evolution, quantum
channels $W^{(t)}_{i,j}$ that act only on the impurity, are applied. We
consider time-dependent $W^{(t)}$ since we allow an arbitrary control signal
applied to the impurity. Thus, density matrix dynamics evolution over one step
reads
\begin{align}
    \rho_{i\alpha}\left(t+\frac{1}{2}\right) &= V_{i\alpha,i'\alpha'} \rho_{i'\alpha'}(t),\nonumber\\
    \rho_{i\alpha}(t+1) &= W^{(t)}_{i,i'} \rho_{i'\alpha}\left(t+\frac{1}{2}\right). \label{eq:tevo}
\end{align}
Integrating Eq.~\eqref{eq:tevo} forward in time and substituting the
integration result into Eq.~\eqref{eq:opdef}, we express the dynamics of the
impurity observable as follows,
\begin{eqnarray}
    \av{O(t)} &&= O_{i_t}\id_{\alpha_t}\left(\prod_{\tau=1}^t W^{(\tau-1)}_{i_\tau j_{\tau-1}}V_{\alpha_\tau j_{\tau-1},\alpha_{\tau-1} i_{\tau-1} }\right)\nonumber\\ &&\times\rho^{\env}_{\alpha_0} \rho^{\imp}_{i_0}.\label{eq:obs_tn}
\end{eqnarray}
It is convenient to
represent the right part of Eq.~\eqref{eq:obs_tn} in terms of a tensor network diagram, as illustrated in Fig.~\ref{fig:ims}a.  Further, we
rewrite Eq.~\eqref{eq:obs_tn} as 
\begin{align}
	\av{O(t)} =  \im_{\{j_\tau,i_\tau\}_{\tau=0}^{t-1}}O_{i_t}\left(\prod_{\tau=0}^{t-1} W^{(\tau)}_{i_{\tau + 1} j_{\tau}}\right) \rho^{\imp}_{i_0},\label{eq:obs_final}
\end{align}
where  the IM $ \im_{\{j_\tau,i_\tau\}_{\tau=0}^{t-1}}$ is defined as follows,
\begin{align}
	\im_{\{j_\tau,i_\tau\}_{\tau=0}^{t-1}} = \id_{\alpha_t}\left(\prod_{\tau=0}^{t-1} V_{\alpha_{\tau+1} j_\tau,\alpha_\tau i_\tau }\right) \rho^{\env}_{\alpha_0}.\label{eq:im}
\end{align}
This IM is a generalization of a Feynman-Vernon influence
functional~\cite{feynman2000Theory}.  In Fig.~\ref{fig:ims}a we highlight IM's
tensor diagram by the shaded area. The IM encapsulates all the effects of the environment on an impurity, including dissipation and memory (non-Markovianity) of the
environment. One advantage of the IM approach is that it allows one to
describe arbitrary local impurity dynamics. In particular, arbitrary control
sequences can be applied by selecting the channels $W^{(\tau)}$ which act on the impurity, since they are not included in the IM.

We can view Eq.~\eqref{eq:im} as an MPS representation of the IM. The quantum
channels $V$ take the role of MPS tensors, and the dimension of the environment
density matrix becomes the bond dimension. This
suggests that conventional MPS compression algorithms such as singular value decomposition-based
truncation~\cite{schollwock2011density} may be used to approximate an IM by an MPS with a lower bond dimension, effectively reducing the  dimensionality of environment.
An IM can be efficiently compressed provided its temporal entanglement -- the
entanglement entropy of the IM viewed as a wavefunction in the time domain --
is moderate. It was found that for a wide variety of physically interesting
cases, TE grows slowly with evolution time. This includes quantum circuits that
are nearly dual unitary~\cite{lerose2021Influence}, many-body localized
systems~\cite{sonner2022Characterizinga}, dissipative
systems~\cite{sonner2021Influence}, free fermion
systems~\cite{thoenniss2023Efficient,thoenniss2023Nonequilibrium}, integrable
systems~\cite{lerose2021Scaling,giudice2022Temporal} and spin-boson
models~\cite{vilkoviskiy2024bound}. Additionally TE generally remains low in
proximity to those regimes~\cite{lerose2021Influence}. Using a compressed, low bond-dimension approximation of an IM,  evolution of local observables can be efficiently computed, see Eq.~\eqref{eq:obs_tn}. 

Below we will be interested in analyzing transport through an impurity
connecting two environments (left and right environments, see
Fig.~\ref{fig:ims}b). For concreteness, we
choose a two-site impurity where the left environment only couples to the left
site of the impurity and the right environment only couples to the right site
of the impurity. The corresponding tensor diagram for observable dynamics in
this model $\av{O(t)}$ is illustrated in Fig.~\ref{fig:ims}b. Note that left
and right IMs enter the tensor network in Fig.~\ref{fig:ims}b independently.
This allows us to construct compressed IMs for the left and right environment
independently and later combine them to build the more complex full
environment. In the same fashion, multiple IMs can be
combined~\cite{fux2023tensor}.

The case where observable $\hat{O}$ corresponds to a conserved charge is of
particular interest for transport properties. To compute the instantaneous
current flowing between left and right parts of the impurity
$I(t)$, we generalize Eq.~\eqref{eq:tevo} and
Eq.~\eqref{eq:opdef} to an impurity coupled to a left and right environment.
The generalization of Eq.~\eqref{eq:tevo} reads
\begin{align}
	\rho_{i k \alpha \beta}\left(t+\frac{1}{2}\right) &= V_{i \alpha,i' \alpha'}^L \rho_{i'k'\alpha'\beta'}(t)V_{k\beta,k'\beta'}^R,\nonumber\\
	\rho_{i k \alpha \beta}(t+1) &= W_{ik,i'k'}^{(t)} \rho_{i' k' \alpha \beta}\left(t+\frac{1}{2}\right),\label{eq:rl_tevo}
\end{align}
where indices $i,\ k$ of $\rho_{ik\alpha\beta}$ correspond to the impurity
degrees of freedom, index $\alpha$ to the left environment degrees of freedom
and index $\beta$ to the right environment degrees of freedom, $V^R$ and $V^L$
are quantum channels forming left and right IMs correspondingly.
The generalization of Eq.~\eqref{eq:opdef} reads
\begin{align}
	\av{O(t)} = O_{i}\id_k\id_{\alpha}\id_{\beta}\rho_{ik\alpha\beta}(t),\label{eq:rl_opdef}
\end{align}
assuming that we consider an observable of the left component of the impurity. Eqs.~(\ref{eq:rl_tevo}),(\ref{eq:rl_opdef}) allow us to calculate a
current $I(t)$ as a finite difference of a charge $O$ before and
after the action of $W$, i.e.
\begin{align}
    I(t) = \av{O\left(t+1\right)} - \av{O\left(t+\frac{1}{2}\right)}.
    \label{eq:fin_diff_current}
\end{align}
For bounded observables and time independent dynamics we expect that the system
reaches a steady state, and the long-time averages of the left and right
current are equal to the steady-state current,
\begin{equation}
\overline{I(t)} = I_s.
\end{equation}
The latter can be calculated as a long time asymptotic of
Eq.~\eqref{eq:fin_diff_current}. Replacing the original IMs by their compressed
MPS representations makes this approach computationally efficient.

The efficiency of the approach described above hinges on obtaining a
compressed, low bond dimension MPS representation of the IM. While low TE
guarantees the existence of such an MPS, finding it for large environments can
be challenging. The known cases where an MPS representation of system's IM can
be efficiently constructed classically include: (\emph{i}) Interacting
one-dimensional chains, via transversal contraction schemes
\cite{banuls2009Matrix,lerose2023overcoming,frias-perez2022Light}; (\emph{ii})
Non-interacting bosons, which allow for explicit construction of a low bond
dimension MPS representation using a finite memory time
cutoff~\cite{strathearn2018efficient} or using auxiliary
bosons~\cite{vilkoviskiy2024bound}; (\emph{iii}) Free fermions where IM can be
expressed in terms of Gaussian integrals which are converted to MPS
form~\cite{thoenniss2023Nonequilibrium,ng2023real}. Despite these advances,
there are important cases, such as interacting models in higher dimensions,
where no algorithms to obtain the compressed IM are known. 

Moreover, the following simple argument shows that the generic problem of
computing the IM from a quantum circuit description of the environment
is classically hard, even if restricted to IMs with low TE. To
illustrate this, let us take an environment which is disconnected from the impurity qubit for the first $t-1$ steps of evolution. The degrees of freedom of the environment evolve according to a quantum circuit encoding a
classically hard quantum computation, with an output written to the state
of one of the environment's qubits. At the last step, the state of this qubit
is swapped with a qubit in the impurity. The corresponding IM is a product
state and hence has zero TE, yet obtaining it requires simulating an arbitrary quantum computation.

\section{Influence matrix reconstruction from quantum measurements}
\label{sec:influence_matrix_recovery}
\begin{figure}[ht]
  \centering
  \includegraphics[width=1.\linewidth]{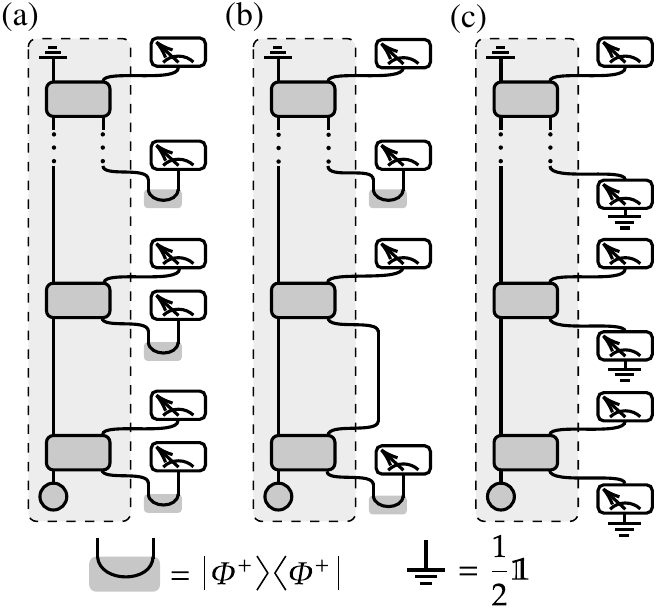}
  \caption{ (a) Illustration of the first protocol of
  data acquisition. At each time step a Bell pair depicted as
  $\ket{\Phi^+}\bra{\Phi^+}$ is prepared, and one of the qubits is coupled to the
  environment. The resulting quantum state, proportional to the
  environment's IM, can be characterized by measuring qubits in different bases. In this scheme, measurements and evolution can be parallelized. (b) Illustration of a coarse grained protocol, where an ancilla interacts with the environment over two time steps. 
  (c) Illustration of the second protocol, which requires one ancilla. At each
  step, the ancilla is prepared in an infinite temperature state depicted as
  $\frac{1}{2}\id$, measured using a SIC-POVM, coupled to the environment before being measured again. This protocol
  provides the same statistics of measurements outcomes. }
  \label{fig:measurement}
\end{figure}

In this Section, we formulate a hybrid quantum-classical approach to learning
the IM, building on previous works on PT tomography~\cite{Guo_2020,White_2022}.
The aim of this approach is to reconstruct a compressed approximation of a
complex environment's IM with moderate TE, in cases where no classical
algorithm is readily available, e.g. for interacting environments in $d\geq 2$
dimensions. 

At the first step, an experiment with a quantum computing device is performed:
a quantum environment $E$ is simulated, probe qubits are brought to interact
with $E$ at different times, and a set of measurements is performed, as
described below. In the second, `classical` stage we use a likelihood
maximization based reconstruction method to convert those measurement results
into a low bond dimension MPS representation of the IM. 

Below we discuss two alternative implementations of the first step, which
differ in the way ancilla qubits are used. Both schemes sample the same
probability distribution and are therefore equivalent from the theory
perspective. However, they have different requirements in terms of the number
of ancilla qubits and the measurement capabilities of the device; thus, one of
the schemes may be better suited for a given experimental platform. 

In the first implementation, multiple ancilla qubits
are used to prepare a quantum state whose density matrix is proportional to the
IM. Effectively, this quantum state is the Choi state~\cite{choi1975completely,jamiolkowski1972linear,pollock2018NonMarkovian} of the IM viewed as a quantum channel between $t$ input and $t$ output states.
This implementation is illustrated in Fig.~\ref{fig:measurement}a: at each step, a Bell pair of ancilla qubits is prepared and one qubit of the pair interacts with $E$ over one time step. We repeat this procedure for each time step, preparing a state of $2t$ ancilla qubits. The density matrix of this state is equal to the IM, up to a normalization constant. Next, a symmetric, informationally complete positive operator-valued (SIC-POVM) measurement on each ancilla qubit is performed (see Appendix~\ref{app:sic-povm} for details).
Each SIC-POVM measurement yields one of the 4 different possible SIC-POVM
elements $\{M^\alpha\}_{\alpha=0}^3$, generating a string of $2t$ integer
indices $\{0, 1, 2, 3\}$ identifying a particular SIC-POVM element. This
`asynchronous' approach requires $2t$ ancilla qubits,  
and the measurement of a given auxiliary pair can be performed at any
moment after the interaction with the environment. It allows one to parallelize evolution and measurements of each qubit -- a potential advantage for platforms where measurements are significantly slower than unitary evolution.

In contrast to the first asynchronous approach, the second, fully sequential approach requires only one ancillary qubit that is measured
multiple times. This implementation is schematically illustrated in Fig.~\ref{fig:measurement}c:
At each step of time evolution, the ancilla qubit is (i) initialized in an
infinite temperature state; (ii) a SIC-POVM on the ancilla is performed; the
qubit is coupled to the environment for one time evolution step, and, finally
(iv) another SIC-POVM is performed on the impurity qubit. After $t$ steps of
time evolution we obtain a measurement string of length $2t$. This approach
does not require multiple ancilla qubits, but one needs to alternate the environment time evolution with
SIC-POVM measurements.

To reconstruct the IM from measurements, the experiment is repeated $N \gg 1$
times. We organize the measurement strings into a dataset matrix $\ds_{k, l}$
where $k$ runs over $N$ strings and $l$ runs over the $2t$ individual SIC-POVM
results in each string. In order to access the information about longer time
scales with fewer samples it may be advantageous to perform measurements after
two or more steps of ancilla-bath interaction. This is illustrated in
Fig.~\ref{fig:measurement}b, where the ancilla is measured after two steps of
interaction with $E$. We use such {\it coarse graining} procedure in some of
our numerical experiments; more information about coarse graining is provided in Appendix~\ref{app:mldetails}.

In the `classical` part of the algorithm we reconstruct a MPS representation of
the IM from the set of measurement strings obtained in the quantum part. As
established above, many relevant quantum environments have low TE and can hence
be described by relatively few degrees of freedom even for a large number of
time steps. To take advantage of this property, we use a low bond dimension
ansatz and ``learn'' its parameters from the measurement results. The bond
dimension is seen as a hyperparameter of the learning algorithm and can be
tuned to regulate the upper bound of TE. Our ansatz is structurally identical
to Eq.~\eqref{eq:im} and reads
\begin{align}
	\im_{\{j_\tau, i_\tau\}_{\tau=0}^{t-1}}(\Theta, \rho) = \id_{\alpha_t}\left(\prod_{\tau=0}^{t-1} \Theta_{\alpha_{\tau+1} j_\tau,\alpha_\tau i_\tau }\right) \rho_{\alpha_0},\label{eq:ansatz}
\end{align}
where $\Theta$ is an unknown quantum channel and $\rho$ is an unknown density
matrix. To find the free parameters $\Theta$ and $\rho$, we maximize a
logarithmic likelihood function of the dataset $\ds_{k,l}$ with respect to
those unknown parameters. The logarithmic likelihood function reads
{\small\begin{align}
		\llh(\ds, \Theta, \rho) &= 	\sum_{k=1}^{N}\log\Bigg(\im_{\{j_\tau,i_\tau\}_{\tau=0}^{t-1}}(\Theta, \rho)\nonumber\\&\times\prod_{\tau=0}^{t-1}M_{j_\tau}^{\ds_{k,2\tau + 2}} M_{i_\tau}^{\ds_{k,2\tau + 1}}\Bigg)- 2tN\log(2).
	\label{eq:llh}
\end{align}}
Here $\{M_i^\alpha\}_{\alpha=0}^3$ is a SIC-POVM with vectorized elements,
where $\alpha$ runs over all elements and $i$ runs over all entries of a
particular element. The last term in the r.-h.s. of Eq.~\eqref{eq:llh}, $2tN\log(2)$, is a
normalization constant that can be safely omitted. The optimization problem
that yields the most likely parameters $\Theta$ and $\rho$ can therefore be summarized as follows,
\begin{align}
	&\underset{\Theta, \rho}{\rm maximize} \quad \llh(\ds, \Theta, \rho)\nonumber\\
	&{\rm subject \ to} \quad \Theta \in {\rm CPTP}, \ {\rm Tr}(\rho)=1, \ \rho \geq 0,
\end{align}
where CPTP stands for ``completely positive and trace preserving'', i.e. the
set of all quantum channels of a fixed dimension. We solve this optimization problem
using automatic differentiation to calculate a gradient with respect to the
unknown parameters and Riemannian optimization \cite{luchnikov2021QGOpt,
luchnikov2021Riemannian} to run a stochastic gradient descent based
optimization procedure preserving constraints. For details on the optimization
procedure and hyperparameters we refer to Appendix~\ref{app:optdetails}.

\section{Results}
\label{sec:results}

In this Section, to evaluate the performance of the hybrid learning procedure
described above, we apply it to reconstruct an IM for two examples of
spin-chain environments. To generate measurement samples, we use an IM that is
computed classically using the algorithm of Ref.~\cite{lerose2023overcoming}.
We find that $10^6-10^7$ measurement strings are sufficient to
accurately reconstruct long-time IMs (precise parameters defined below). The
reconstructed IMs can be used to compute quantum impurity
observables, including transport in setups with more than one environment. 

We consider an environment that is a semi-infinite chain of spinless fermions
with nearest-neighbour interactions. Performing Jordan-Wigner transformation,
such environment is mapped onto an XXZ spin chain. Discretizing time, the
Hamiltonian evolution can be approximated by repeatedly applying the following
Floquet operator, 
\begin{equation}
    U(J,J')=U_\mathrm{e}(J,J') U_\mathrm{o}(J,J'),
\end{equation}
\begin{align}
	&U_\mathrm{e}\left(J,J'\right)=e^{-i \sum_{i=0}^\infty H_{2i,2i + 1}\left(J, J'\right)},\nonumber\\
	&U_\mathrm{o}\left(J,J'\right)=e^{-i \sum_{i=0}^\infty H_{2i+1,2i+2}\left(J, J'\right)},\nonumber\\
    &H_{i, j}\left(J, J'\right) = J \left(X_{i} X_{j} + Y_{i} Y_{j}\right) + J' Z_{i} Z_{j},
	\label{eq:result_model}
\end{align}
where a subscript $\mathrm{e(o)}$ stands for even (odd), and $X_i,Y_i,Z_i$ are
the Pauli matrices acting on the $i$-th qubit. Effectively, the evolution is
therefore represented by a brickwork quantum circuit. 

Below we will focus on two particular parameter choices: the free-fermion case,
which corresponds to $J'=0$ (the XX-model), and the Heisenberg point, $J'=J$
(the XXX-model). The XX-model exhibits an area-law
TE~\cite{giudice2022Temporal} for a broad class of initial states, which
includes thermal states at any non-zero finite temperature. Furthermore, XX model has a
ballistic spin transport. The XXX model displays TE that grows logarithmically
in time~\cite{giudice2022Temporal}, and exhibits anomalous spin transport,
including superdiffusion in the linear-response
regime~\cite{ljubotina2019kardar}. We consider two values of the
interaction parameter $J$, $J=0.1$ and $J=0.2$. These values are sufficiently
small to approximate the Hamiltonian dynamics without appreciable heating due
to absence of energy conservation in the Floquet evolution. The number of time
steps is fixed at $t=60$. Further, we consider 3 initial states of the
spin-chain environment: a fully mixed infinite temperature state as well as two
fully polarized states, defined by the density matrices:
{\small\begin{align}
	\rho_{\infty} =
	\bigotimes^{\infty}_{i=1} \frac{1}{2}\id_i, \ \, \rho_{\uparrow} =\bigotimes^{\infty}_{i=0}\ket{\uparrow_i}\bra{\uparrow_i}, \ \, \ \rho_{\downarrow} = \bigotimes^{\infty}_{i=0}\ket{\downarrow_i}\bra{\downarrow_i}.
\end{align}}
All three states are stationary states of environment's Hamiltonian. With these
initial states, the spin chains exhibit a moderate TE, which allows for an efficient compression of their IM as MPS. 

We first classically compute the ``first-principles" IM $\im_{\rm f}$ using a
light-cone growth algorithm described in Ref.~\cite{lerose2023overcoming}, with
a maximum bond dimension of $\chi=256$. We verified that the obtained IMs are
well-converged with respect to the bond dimension. Further, for each IM we
generate a dataset $\ds_{k,l}$ consisting of $N=0.2\cdot 10^6$ to $N=5\cdot
10^6$ measurement strings using perfect sampling from the MPS form of the IM,
mimicking a probability distribution of the measurement
outcomes~\cite{ferris2012perfect, chertkov2022optimization}. We then use the
procedure outlined in Section~\ref{sec:influence_matrix_recovery} to
reconstruct the IMs from the measurement results datasets $\ds_{k,l}$. For details regarding the learning process and
hyperparameters, see Appendix~\ref{app:mldetails}.

\begin{figure}
	\centering
	\includegraphics[scale=0.75]{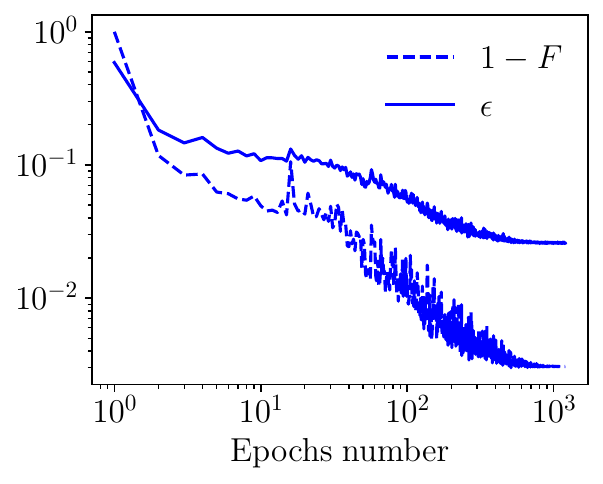}
	\caption{Comparison of the infidelity $1-F$ and the prediction error $\epsilon$
	against the number of learning epochs during the learning process of the IM
	for the XXX model. The XXX-model is defined by a Floquet operator in
	Eq.~\eqref{eq:result_model} with $J=J'=0.1$. The initial state of the system	and environment is $\ket{\uparrow}\bra{\uparrow}\otimes\rho_{\downarrow}$, the dataset size is $N = 5\cdot 10^6$. Both metrics show that the IM can be
	reconstructed faithfully. }
	\label{fig:tracking}
\end{figure}

\begin{figure}
    \centering
    \includegraphics[scale=0.75]{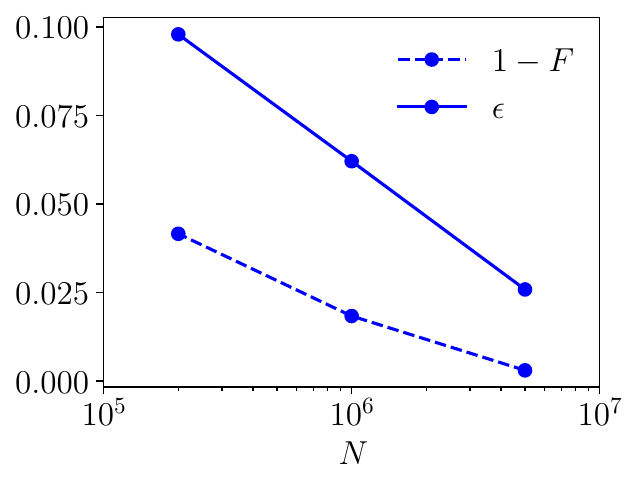}
    \caption{Comparison of the infidelity $1-F$ and the
    prediction error $\epsilon$ at the final point of the learning process
    against the dataset size for the XXX model. Both metrics show that quality
    of IM reconstruction and prediction accuracy systematically improve with increasing the dataset size.}
    \label{fig:accuracy_vs_dataset_size}
\end{figure}

We track the learning process by computing the prediction error $\epsilon$ as
well as infidelity $1-F$ of the reconstructed IM after each complete traversal
of a dataset matrix $\mathbf{X}_{k, l}$ also referred to as a learning epoch.
The prediction error $\epsilon$ is computed as follows
\begin{eqnarray}
	\epsilon = \overline{\frac{1}{t}\sum_{\tau=1}^t\left\|\rho^{\imp}_{\rm f}(\tau) - \rho^{\imp}_{\rm r}(\tau)\right\|_1},
\end{eqnarray}
where $\rho^{\imp}_{\rm f}(\tau)$ is the dynamics of the impurity coupled with
the first-principles IM and $\rho^{\imp}_{\rm r}(\tau)$ is the dynamics of the
impurity coupled with the reconstructed IM and $\|\cdot\|_1$ stands for the
trace norm. A horizontal bar denotes averaging over $4000$ sequences of random
unitary channels $\{W^{(0)},\dots W^{(t-1)}\}$  acting on the impurity in between interactions
with environment (see Fig.~\ref{fig:ims}). The infidelity $1-F$ is calculated
as if IM were a wavefunction of a pure quantum state, i.e.
\begin{align}
	1-F= 1-\frac{|\langle\im_{\rm f}|\im_{\rm r}\rangle|^2}{\langle \im_{\rm r}|\im_{\rm r} \rangle \langle \im_{\rm f}|\im_{\rm f}\rangle},
\end{align}
where $\im_{\rm f}$ is the ``first-principles" IM and $\im_{\rm r}$ is the
corresponding reconstructed IM. We note that the impurity dynamics and the
overlaps can be computed efficiently using the MPS form of the IMs. These two
metrics for the case of XXX model with $J=0.1$ are plotted across the learning
process in Fig.~\ref{fig:tracking}, while their values at the end of the
learning process are plotted against the dataset size in Fig.
\ref{fig:accuracy_vs_dataset_size}.

During the learning process the infidelity quickly drops below $10^{-2}$ for
$N=5\cdot 10^6$ samples. This means that the relative error of the IM
reconstruction is small for machine learning approaches. However, in our
case the IM is a large tensor with $2^{4t}$ elements, and its norm strongly depends on the nature of the environment. Specifically, its Frobenius norm
satisfies an inequality $1\leq\sqrt{\langle \im_{\rm r}|\im_{\rm r} \rangle}\leq 2^t$,
where the lower bound is saturated when IM is a fully depolarising quantum channel, and
the upper bound is saturated for an IM which is seen as a unitary
quantum channel. Thus, Frobenius norm of $\ket{\im_{\rm r}}$ could be
exponentially large in general.
Since an expectation value of an impurity observable is proportional to $\ket{\im_{\rm r}}$ without the normalization factor,
its absolute error could be as large as $\mathcal{O}\left(\sqrt{\left(1 -
\sqrt{F}\right)\langle \im_{\rm r}|\im_{\rm r} \rangle}\right)$. Therefore, 
even a small relative error could in principle significantly affect the prediction
accuracy for impurity observables. However, we find that this is not the case:
the prediction error $\epsilon$ reaches values of $\epsilon\approx 0.025$ for
$N=5\cdot 10^6$ at the end of the learning protocol, indicating that the impurity dynamics is reproduced faithfully. Both metrics become stationary after
around $10^3$ epochs, indicating that the learning process is converged. 

The above results indicate that the reconstructed IM allows an accurate
prediction of the dynamics of an impurity coupled to a single environment.
Furthermore, the IMs learned in an experiment with a single environment has a
broader applicability. In particular, it can be used to make predictions in
quantum transport setups that involve multiple environments (leads), for
arbitrary time-dependent impurity Hamiltonian. 

We illustrate this by studying time evolution of an impurity coupled to left
and right environments, which are chosen to be semi-infinite XX or XXX chains,
with initial states $\rho^{L(R)}$ and corresponding IMs $\im^{L(R)}$. We choose
the impurity to consist of two spins coupled by a generic XYZ interaction; it
is then convenient to think of the system as a spin chain, with the impurity
sites at $i=0,1$. The quantum channel applied to the impurity over one time
step is given by,  
\begin{align}
	&W(J_x, J_y, J_z)[\cdot] = U_{\rm imp}(J_x,J_y,J_z)\cdot U_{\rm imp}^\dagger(J_x,J_y,J_z),\nonumber\\
	&U_{\rm imp}(J_x,J_y,J_z) = e^{-i\left(J_x X_0 X_1 + J_y Y_0 Y_1 + J_z Z_0 Z_1\right)},
	\label{eq:impurity}
\end{align} 
where $\cdot$ denotes the density matrix of impurity. Note that the choice of
parameters $J_x = J_y = J$ and $J_z = J'$ corresponds to a homogeneous infinite
chain. 

\begin{figure}
	\centering
	\includegraphics[scale=.75]{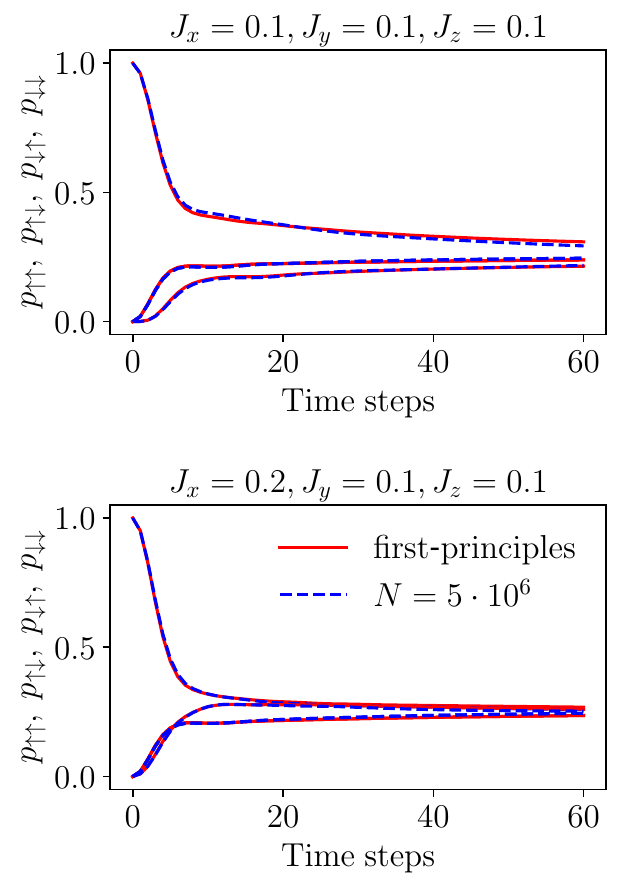}
	\caption{Dynamics of diagonal elements of the impurity's density matrix, $p_{\uparrow\uparrow}=\bra{\uparrow}\otimes\bra{\uparrow}\rho^{\imp}\ket{\uparrow}\otimes\ket{\uparrow}$, $p_{\uparrow\downarrow}=\bra{\uparrow}\otimes\bra{\downarrow}\rho^{\imp}\ket{\uparrow}\otimes\ket{\downarrow}$, $p_{\downarrow\uparrow}=\bra{\downarrow}\otimes\bra{\uparrow}\rho^{\imp}\ket{\downarrow}\otimes\ket{\uparrow}$, $p_{\downarrow\downarrow}=\bra{\downarrow}\otimes\bra{\downarrow}\rho^{\imp}\ket{\downarrow}\otimes\ket{\downarrow}$,
	for the transport setup with environments being XXX semi-infinite spin chains, defined in
	Eq.~\eqref{eq:result_model}, with $J=J'=0.1$. Impurity evolution is given by
	Eq.~\eqref{eq:impurity}. The solid red curves represent results
	obtained using first-principles IMs, while the dashed blue curves are obtained
	from reconstructed IMs. The dataset size used to
	reconstruct IMs is $N = 5\cdot 10^6$.} \label{fig:xxx_relaxation}
\end{figure}

We start by considering a quantum transport setup where impurity is initialized
in an out-of-equilibrium state and is coupled to the leads at time $t=0$. For
this numerical experiment, the initial density matrices of the left and right
environments are chosen to be infinite-temperature states. We set the initial
density matrix of the impurity to the following pure state
\begin{align}
	\rho^\imp(0) =\ket{\uparrow}\bra{\uparrow}\otimes\ket{\uparrow}\bra{\uparrow},
\end{align}
and $J=J'=0.1$ for both environments. In Fig.~\ref{fig:xxx_relaxation} we
compare the dynamics of the diagonal elements of $\rho^I(t)$ computed using the
first-principles IM with the dynamics computed with the IM reconstructed using
a dataset of size $N=5\cdot 10^6$. As impurity's quantum channel parameters we
choose  $J_x = J_y = J_z = 0.1$ corresponding to the integrable homogeneous
chain and $J_x = 0.2$, $J_y= J_z = 0.1$ corresponding to a non-integrable
chain with an impurity.
It is evident that the time evolution obtained using the reconstructed IMs
closely follows that obtained from first-principles IMs. This is a highly
nontrivial observation, since two environments interact with each other via the
impurity and the physics of this interaction is correctly captured by the IMs
reconstructed from a numerical experiment with a single environment.

\begin{figure}[ht]
	\centering
	\includegraphics[scale=0.75]{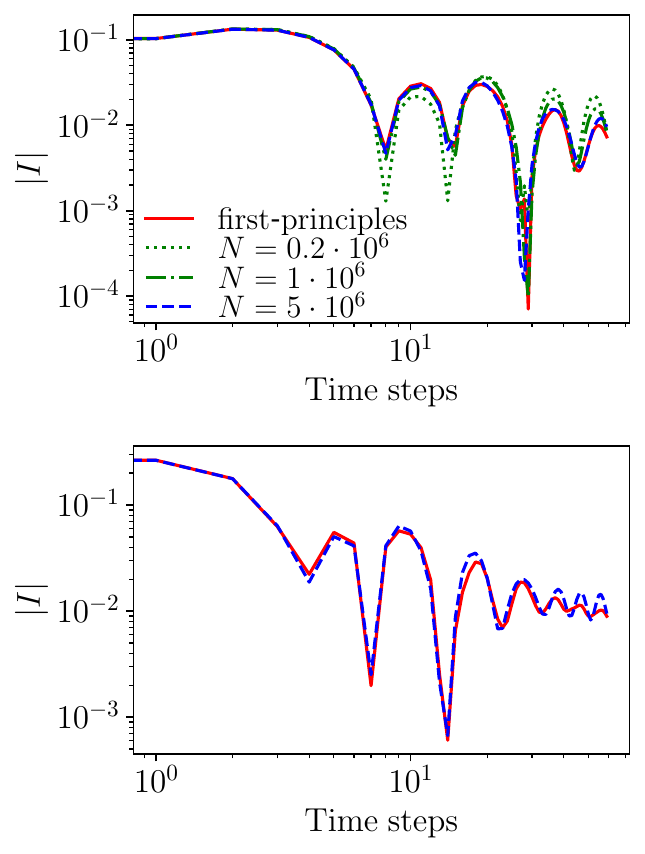}
	\caption{Instantaneous current flowing through the impurity coupled with two
	XXX environments Eq.~\eqref{eq:result_model} for $J=J'=0.1$ (top panel)
	and for $J=J'=0.2$ (bottom panel). Solid red curves are obtained by using
	first-principles IMs, dashed curves are obtained by using the reconstructed IMs
	from datasets of different size.}
	\label{fig:xxx_current_dynamics}
\end{figure}

Next, we consider relaxation of an initial domain wall configuration in a
homogeneous XXX spin chain, corresponding to the choice of parameters, $J_x =
J_y = J$ and $J_z = J'=J$. In this case, the total $Z$-projection of spin is
conserved. The initial state of the left environment is chosen to be polarized
down $\rho^L = \rho_\downarrow$ and the initial state of the right environment
to be polarized up $\rho^R = \rho_\uparrow$,  while an impurity is initialized
in a state
\begin{align}
\rho^\imp(0) =
\ket{\downarrow}\bra{\downarrow}\otimes\ket{\uparrow}\bra{\uparrow}, 
\end{align}
corresponding to a domain wall at $x=1/2$. Similar setups have been recently
probed experimentally \cite{wei2022quantum,rosenberg2023dynamics}. Setting
$J=J'=0.1$ and $J=J'=0.2$, we computed the instantaneous current through the
domain wall by  Eq.~\eqref{eq:fin_diff_current} using first-principles IM and
the reconstructed ones. The comparison of these two computations for different
dataset sizes $N$ is  provided in Fig.~\ref{fig:xxx_current_dynamics}. One
observes that the current dynamics prediction based on the reconstructed IMs
improves with increasing dataset size. While for short times the agreement is
near perfect, the results based on the reconstructed IMs exhibit slight
deviation from the reference ones even for the largest dataset size, $N=5\cdot
10^6$. 

\begin{figure}[ht]
	\centering
	\includegraphics[scale=0.75]{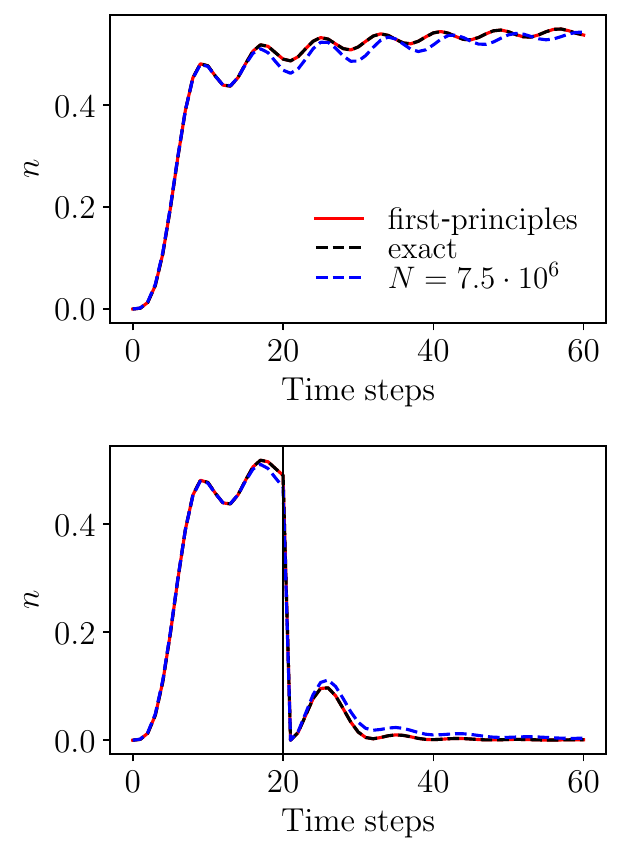}
	\caption{Dynamics of the occupation number on the impurity's left qubit
	under a control signal in Eq.~\eqref{eq:control_seq} (bottom panel) and without	time dependent control (top panel). Solid red curves are computed by using the first-principles IM, dashed black curves are computed exactly using the
	free fermion representation, and dashed blue curves are computed with use of
	the reconstructed IM. The vertical black line in the bottom panel highlights the moment when	particles are removed from the impurity and two sites of the impurity are	decoupled.}
	\label{fig:xx_memory_probing}
\end{figure}

Finally, we demonstrate that reconstructed IMs correctly capture
long-time non-Markovian effects, as well as a response to an external
control signal. To that end, we consider a two-site impurity in an initial state
$\rho^\imp(0) =
\ket{\downarrow}\bra{\downarrow}\otimes\ket{\downarrow}\bra{\downarrow}$
coupled with two XX semi-infinite chains with $J = 0.1, J'=0$ and initial states $\rho^L =
\rho_\downarrow$, $\rho^R = \rho_\uparrow$. The time-dependent  impurity channel $W^{(\tau)}$ is chosen as follows
\begin{align}
	W^{(\tau)}[\cdot] = \begin{cases}
		U_{\rm imp}(J, J, J') \cdot U^\dagger_{\rm imp}(J, J, J'), \ \tau < 20,\\
		\mathcal{D}[\cdot], \ \tau = 20,\\
		{\rm Id}[\cdot], \ \tau > 20,
	\end{cases}
	\label{eq:control_seq}
\end{align}
where $\mathcal{D}$ is an amplitude damping channel turning any impurity's
state into $\ket{\downarrow}\bra{\downarrow}\otimes
\ket{\downarrow}\bra{\downarrow}$, and ${\rm Id}$ is the identity channel. Note
that the impurity dynamics in this case can be calculated exactly, since this
model is non-interacting. Impurity dynamics described by Eq.~\eqref{eq:control_seq} can be described as
follows: first, at times $t<20$, there is a current flowing from right to left part of the system, and density on the impurity is increasing towards a steady-state value of $n_0=1/2$. Further, at time $t=20$, the impurity is reset to a state without particles, and left and right sites of the impurity are decoupled, such that each site is now coupled only to one environment. 
During subsequent evolution, we expect the occupation number of the left impurity site, $n(\tau)$ to exhibit a peak due to the backflow of particles which were transferred from the right environment to the left one during time $t<20$, before settling to a value of $n=0$ at long times. Such a peak is a reflection of non-Markovianity of system's dynamics. In 
Fig.~\ref{fig:xx_memory_probing} we compare dynamics of the particles number on the left impurity qubit ${n}(\tau)$ computed exactly, with the first-principles IM and with the reconstructed IM. 
The reconstructed IM follows the exact results closely, in particular reproducing the peak around the time $t=25$. We provide additional numerical results for other environments and impurity time  evolution in Appendix~\ref{app:xx_density_dynamics}.

\section{Discussion and Outlook}\label{sec:Outlook}



In summary, in this paper we investigated a hybrid algorithm for
reconstructing IMs of large many-body environments. Using several examples,
where measurement data were mimicked using a classical  computation, we found
that the IMs for long-time evolution could be efficiently learned from a
relatively limited number of measurements. The efficiency of this approach
relies  on the moderate scaling of IMs temporal entanglement with evolution
time,  which allows for an MPS representation. Our work builds on, and
complements previous works, Refs.~\cite{white2020demonstration,White_2022}, where a
process tensor was reconstructed from experimental data, for a few time steps. 

The MPS representation of an environment's IM allows for efficient classical
computations of arbitrary dynamics of a small quantum system interacting with
that environment.  In particular, as an application of the proposed algorithm,
we considered several quantum transport setups in 1d systems, including an
impurity coupled to XX and XXX spin half-chain  environments, as well as a
quantum quench starting from a domain wall state. We demonstrated that an IM
learned from an experiment with a {\it single} environment allowed us to
accurately model setups with multiple leads. We envision this to be useful,
e.g. in cases where a quantum processor has enough degrees of freedom to
simulate a single environment, but is not suitable for directly modeling
multiple leads. 

Among the most promising applications of the proposed approach is learning the
structure of IMs for many-body environments where no efficient classical
algorithm is currently available, but there are physical arguments suggesting
the  area-law scaling of the temporal entanglement. One class of such systems
are non-integrable systems with long-lived quasiparticles, e.g. interacting 2d
Fermi liquids. Furthermore, IM learning may shed  light on the issue of TE
scaling in thermalizing many-body systems.  On the one hand, intuitively one
may expect that a large thermalizing systems act as baths with a finite memory
time (thermalization time), leading to an area-law TE;  on the other hand,
recent studies of temporal entanglement in quantum circuits~\cite{lerose2021Influence,FolignoBertiniPRX2023}, in particular in
proximity to dual-unitary points and in random circuits, point to a volume-law
scaling. It would be interesting to apply the approach of this paper to learn IMs for a broader class of thermalizing environments realized using quantum processors. The ability to reconstruct the IM in an MPS form would signal area-law TE. While we generally expect the learning procedure to fail for IMs with volume-law TE, it is an interesting open question whether in such cases there exists an area-law temporally-entangled IM which approximates the environment effect for certain restricted classes of impurity's dynamics.

\section{Acknowledgements}
The computations were performed at University of Geneva using Baobab HPC
service. We thank Alessio Lerose, Julian Thoenniss, and Ilya Vilkoviskiy for
helpful discussions and collaboration on related topics, and Juan Carrasquilla
for insightful discussions. This work was partially supported by the European
Research Council (ERC) under the European Union's Horizon 2020 research and
innovation program (grant agreement No. 864597) and by the Swiss National
Science Foundation.

\bibliography{measure.bib}

\onecolumngrid
\appendix

\section{Vectorization}
\label{app:vectorization}
To introduce the vectorized operators extensively used in the main text, such as in Eq.~\eqref{eq:opdef}, we need to define a basis of operators and enumerate its elements using a single index. For this purpose, we employ a basis of matrix units, which, for a single-qubit case, reads
\begin{eqnarray}
    e_0 = \ket{0}\bra{0}, \ e_1 = \ket{1}\bra{0}, \ e_2 = \ket{0}\bra{1}, \ e_3 = \ket{1}\bra{1}.
\end{eqnarray}
Then, any operator can be vectorized as follows
\begin{eqnarray}
	O_k = \tr\left({e_kO}\right),
\end{eqnarray}
where $O$ is an arbitrary one-qubit operator.
Also, any $m$-subsystem density matrix can be vectorized as follows
\begin{eqnarray}
	\rho_{k_1\dots k_m} = \tr\left(\left[\bigotimes_{i=1}^m e_{k_i}\right]\rho\right).
\end{eqnarray} 

\section{Information complete measurements}
\label{app:sic-povm}
To extract information from the Choi state of an IM, we perform a
symmetric, informationally complete (SIC) positive operator-valued (POVM) measurement of each qubit of the Choi state. A POVM measurement is the most general measurement allowed by quantum theory. It is defined by a set of positive semi-definite Hermitian matrices $\{M^\alpha\}_{\alpha=0}^3$, where each matrix corresponds to a particular measurement result, and these matrices sum to the identity matrix, i.e. $\sum_{\alpha=0}^{3} M^\alpha = \id$. The probability of getting a particular measurement result  is given by the Born rule $p[\alpha] = \tr(\rho M^\alpha)$, where $\rho$ is a density matrix. In the vectorized form it reads $p[\alpha] = M^\alpha_i \rho_i$, where $i$ runs over all entries of a POVM element and a density matrix. Information completeness means that the measured state can be exactly reconstructed from measurement results in the limit of an infinite number of measurements or that the matrix $M^\alpha_i$ is invertible. Symmetry means that the number of POVM elements is the minimal possible guaranteeing information completeness ($4$ for a qubit) and that the POVM is highly symmetric in some sense~\cite{renes2004symmetric}. We choose a particular example of SIC-POVM that is defined
by a set of vectorized operators enumerated by $\alpha \in \{0, 1, 2, 3\}$
\begin{align}
	M^\alpha_{i} = \frac{1}{4}\left(\id_{i} + s^\alpha_k\sigma^k_{i}\right),
\end{align}
where index $i$ runs over all entries of the vectorized POVM element, $\sigma^k_{i}$ are vectorized Pauli matrices enumerated by $k$ and
\begin{align}
	s^\alpha_k = \begin{bmatrix}
		0 & 0 & 1 \\
		\frac{2\sqrt{2}}{2} & 0 & -\frac{1}{3} \\
		-\frac{\sqrt{2}}{3} & \sqrt{\frac{2}{3}} & -\frac{1}{3} \\
		-\frac{\sqrt{2}}{3} & -\sqrt{\frac{2}{3}} & -\frac{1}{3}
	\end{bmatrix},
	\label{eq:tetrahedral_nodes}
\end{align}
where $\alpha$ enumerates the rows and $k$ enumerates the columns. Note, that the rows of
the matrix Eq.~\eqref{eq:tetrahedral_nodes} are the coordinates of the vertices of a tetrahedron, which justifies the symmetry of this POVM. Since each qubit measurement
ends up as a particular value of $\alpha$, the measurement result of the entire
state is a string with numbers from a set $\{0, 1, 2, 3\}$.  A set of these
strings form a dataset that is used for further IM recovery.

\section{Optimization algorithm}
\label{app:optdetails}
The central computational problem of the present algorithm is the following
constrained optimization problem
\begin{align}
	&\underset{\Theta, \rho}{\rm maximize} \quad \llh(\ds, \Theta, \rho)\nonumber\\
	&{\rm subject \ to} \quad \Theta \in {\rm CPTP}, \ {\rm Tr}(\rho)=1, \ \rho \geq 0.
\end{align}
First, we note that any density matrix can be represented as follows $\rho =
\phi[1]$, where $\phi$ is a CPTP map that maps a $1 \times 1$ density matrix (a
scalar $1$) to $\rho$. This brings us to an optimization problem with only CPTP
constraints
\begin{align}
	&\underset{\Theta, \phi}{\rm maximize} \quad \llh^{'}(\ds, \Theta, \phi)\nonumber\\
	&{\rm subject \ to} \quad \Theta, \phi \in {\rm CPTP}.
\end{align}
Note that due to the Stinespring dilation, one can parametrize both CPTP maps taking part in the optimization problem as follows
\begin{align}
	\Theta[\cdot] &= \tr_A\left(V\cdot V^\dagger\right),\nonumber\\
	\phi[\cdot] &= \tr_A\left(v\cdot v^\dagger\right),
\end{align}
where $\tr_A$ is a partial trace over an $r$-dimensional auxiliary space,  $V$
is an isometric matrix of size $rn\times n$, $v$ is an isometric matrix of size
$rn \times 1$, $n$ is a square root of the ansatz bond dimension. Note, that to
parametrize an arbitrary CPTP map one needs $r\geq n^2$ for $V$ and $r\geq n$
for $v$, but taking $r$ smaller we can regularize the learning procedure by
restricting the amount of dissipation introduced by an ansatz. We use $r$ as
another hyperparameter and call it a {\it local Choi rank}. Using the
parametrization above, we turn the optimization problem to the following one
\begin{align}
	&\underset{V, v}{\rm maximize} \quad \llh^{''}(\ds, V, v)\nonumber\\
	&{\rm subject \ to} \quad V\in{\rm St}(nr, n),\ v\in{\rm St}(rn, 1),
\end{align}
where ${\rm St}$ denotes a set of all isometric matrices of a given size also
called a {\it Stiefel manifold}. Since this set is a differentiable manifold,
one can solve the given problem using Riemannian optimization techniques, i.e.,
a variation of gradient descent that operates within a differentiable
manifold. Here we use so-called Riemannian ADAM optimizer which is the
Riemannian generalization of the ADAM optimizer popular in the field of deep
learning.

\section{Details on learning experiments}
\label{app:mldetails}
In our numerical experiments, we considered $8$ different first-principles IMs
that we mark as follows: (\emph{i}) XXX, $J=0.1$, $\rho_{\uparrow}$;
(\emph{ii}) XXX, $J=0.1$, $\rho_{\downarrow}$; (\emph{iii}) XXX, $J=0.1$,
$\rho_{\infty}$; (\emph{iv}) XXX, $J=0.2$, $\rho_\uparrow$; (\emph{v}) XXX,
$J=0.2$, $\rho_\downarrow$; (\emph{vi}) XX, $J=0.1$, $\rho_\uparrow$;
(\emph{vii}) XX, $J=0.1$, $\rho_\downarrow$; (\emph{viii}) XX, $J=0.1$,
$\rho_\infty$. Here XXX or XX classifies a model type, $J$ is a physical time
spent per discrete time step, $\rho_{\uparrow}$, $\rho_{\downarrow}$,
$\rho_{\infty}$ are initial states of an environment.  All IMs
have $60$ discrete time steps and maximum bond dimension of $\chi = 256$. 

For each IM we generated datasets of measurement strings: $3$
datasets of different size for IMs (\emph{i}) and (\emph{ii}), and
one dataset per IM for the rest.  Dataset sizes are given in
the Table~\ref{table:learning_params}.  Some of the datasets were collected using coarse graining: half of each of those datasets was
collected with passing a Bell state's part through $10$ time steps and another
half was collected without coarse graining. Datasets with coarse graining are
emphasized in the Table~\ref{table:learning_params}.

Each dataset was used to reconstruct an IM using the developed
learning algorithm. The bond dimension and the local Choi rank (see
Appendix~\ref{app:optdetails} for a reference on what we call a local Choi
rank) of an ansatz for each case are given in Table~\ref{table:learning_params}.
For all learning experiments we set batch size to $5000$ measurement strings,
initial learning rate to $0.25$. The number of learning epochs and the final
learning rate were different for XXX and XX cases. Their values are given in
the Table~\ref{table:learning_params}.  The learning rate decayed
exponentially from the initial value to the final value during the training. In
all experiments we used Riemannian ADAM optimizer on the Stiefel manifold with
parameters $\beta_1=0.9$, $\beta_2=0.999$, $\epsilon=10^{-8}$.
\begin{table}[ht]
	\caption{Some parameters of learning experiments that vary from one IM to another.}
	\label{table:learning_params}
	\begin{tabular}{|c|c|c|c|c|c|c|c|c|}
		\hline
		& \makecell{XXX,\\$J=0.1$,\\$\rho_\uparrow$} & \makecell{XXX,\\$J=0.1$,\\$\rho_\downarrow$} &
		\makecell{XXX,\\$J=0.1$,\\$\rho_\infty$} & \makecell{XXX,\\$J=0.2$,\\$\rho_\uparrow$} & \makecell{XXX,\\$J=0.2$,\\$\rho_\downarrow$} & \makecell{XX,\\$J=0.1$,\\$\rho_\uparrow$} & \makecell{XX,\\$J=0.1$,\\$\rho_\downarrow$} & \makecell{XX,\\$J=0.1$,\\$\rho_\infty$}\\
		\hline
		\makecell{Dataset size} & \makecell{$0.2\cdot 10^6$ \\ $1\cdot10^6$ \\ $5\cdot 10^6$} & \makecell{$0.2\cdot 10^6$ \\ $1\cdot10^6$ \\ $5\cdot 10^6$} & $5\cdot 10^6$ &  $5\cdot 10^6$ & $5\cdot 10^6$ & $7.5\cdot10^6$ & $7.5\cdot10^6$ & $7.5\cdot10^6$\\
		\hline
		\makecell{Ansatz local\\Choi rank} & \makecell{100} & \makecell{100} & 100 & 100 & 100 & 16 & 16 & 16\\
		\hline
		\makecell{Coarse\\graining} & Yes  & Yes & Yes & Yes & Yes  & No & No & No\\ 
		\hline
		\makecell{Training epochs\\number} & 1200 & 1200 & 1200 & 1200 & 1200 & 2400 & 2400 & 2400\\
		\hline
		\makecell{Final learning\\rate} & $10^{-5}$ & $10^{-5}$ & $10^{-5}$ & $10^{-5}$ & $10^{-5}$  & $10^{-3}$ & $10^{-3}$ & $10^{-3}$\\
		\hline
	\end{tabular}
\end{table}

\section{More numerical results}
\label{app:xx_density_dynamics}
\begin{figure}[ht]
	\centering
	\begin{minipage}{.45\textwidth}
		\includegraphics[width=\columnwidth]{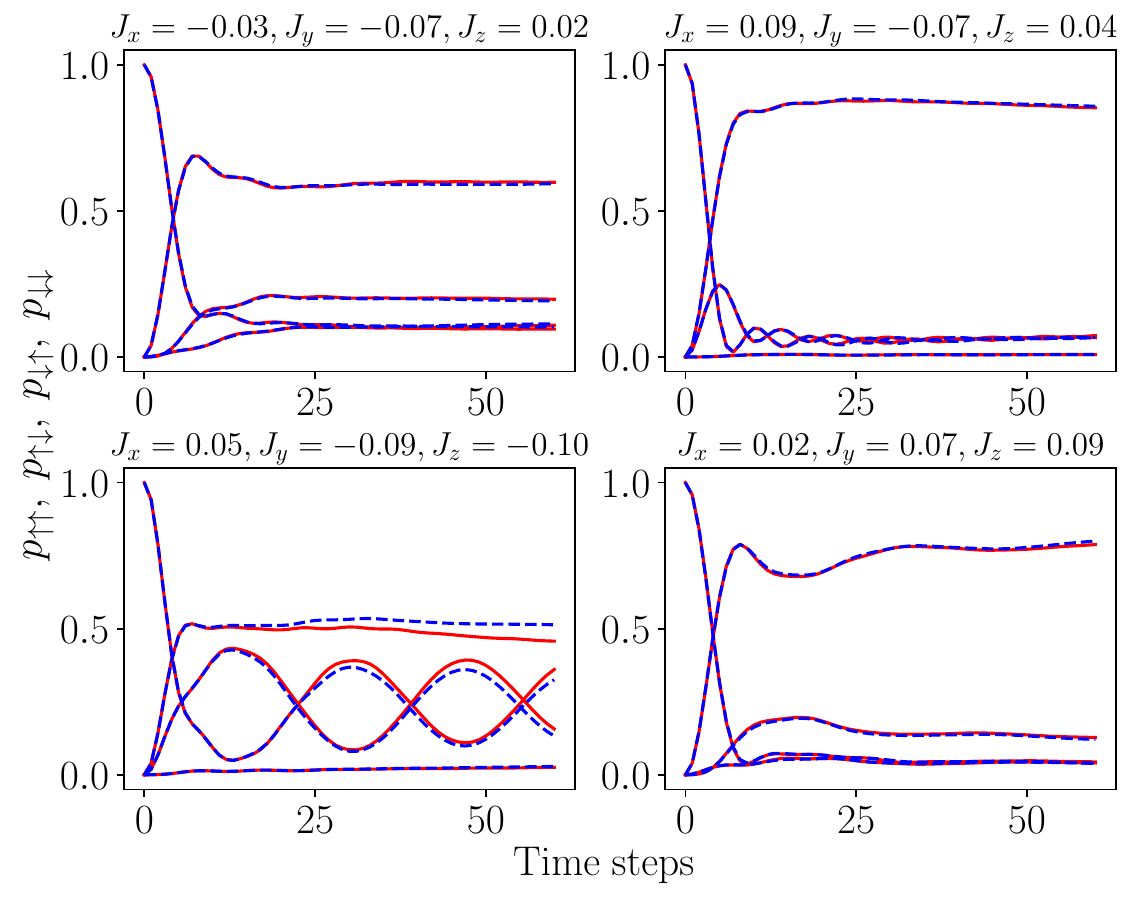}
	\end{minipage}
	\hspace{.05\columnwidth}
	\begin{minipage}{.45\textwidth}
		\includegraphics[width=\columnwidth]{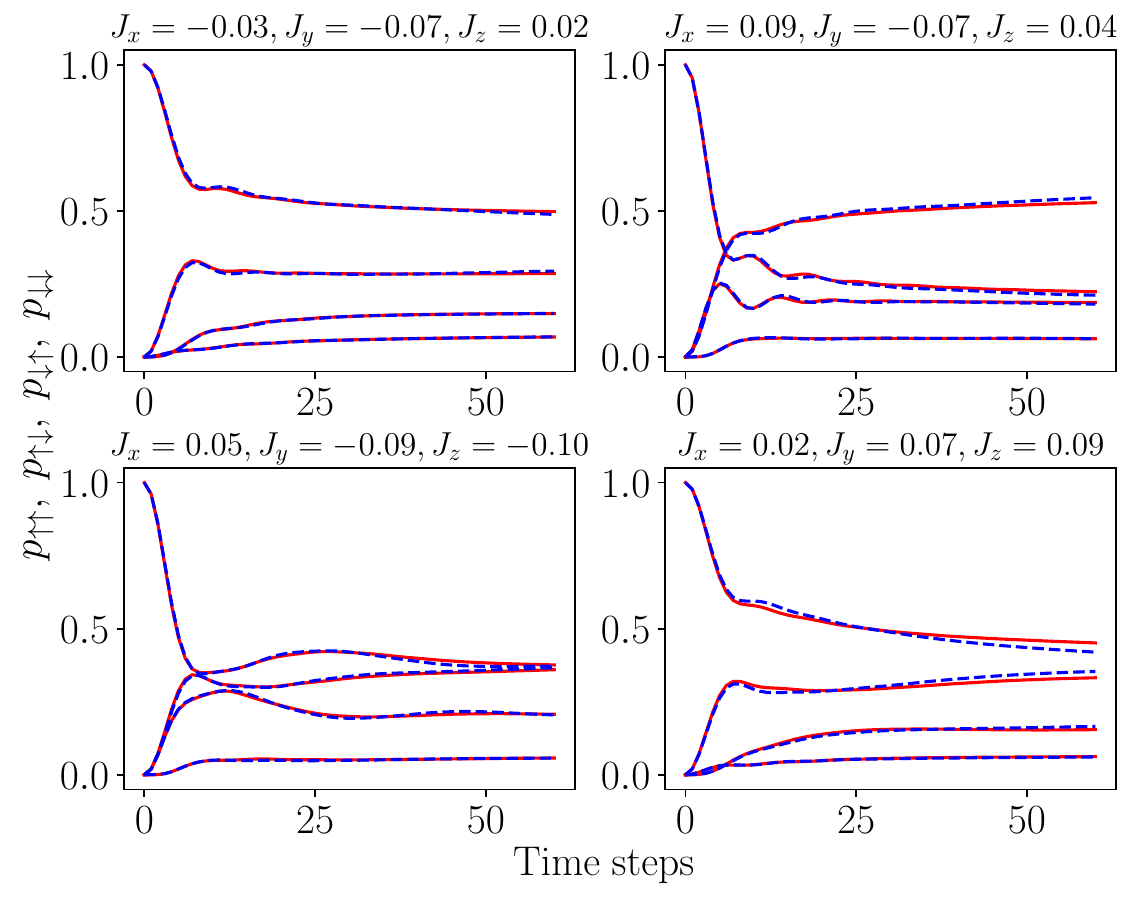}
	\end{minipage}
	\caption{The comparison of the first-principles dynamics (solid lines) of the
	impurity density matrix diagonal elements, that are given by
	$p_{\uparrow\uparrow}=\bra{\uparrow}\otimes\bra{\uparrow}\rho^{\imp}\ket{\uparrow}\otimes\ket{\uparrow}$,
	$p_{\uparrow\downarrow}=\bra{\uparrow}\otimes\bra{\downarrow}\rho^{\imp}\ket{\uparrow}\otimes\ket{\downarrow}$,
	$p_{\downarrow\uparrow}=\bra{\downarrow}\otimes\bra{\uparrow}\rho^{\imp}\ket{\downarrow}\otimes\ket{\uparrow}$,
	$p_{\downarrow\downarrow}=\bra{\downarrow}\otimes\bra{\downarrow}\rho^{\imp}\ket{\downarrow}\otimes\ket{\downarrow}$,
	with the dynamics predicted using learned IMs (dashed
	lines). The left panel corresponds to XXX environments with initial states being $\rho^L = \rho_{\downarrow}$ and $\rho^R = \rho_{\uparrow}$, the right panel corresponds to XXX environments with initial states being $\rho^L = \rho_{\infty}$ and $\rho^R = \rho_{\uparrow}$. The size of datasets used to learn influence matrices is $N=5\cdot
	10^6$. The initial state of the impurity is $\rho^\imp(0) =
	\ket{\uparrow}\bra{\uparrow} \otimes \ket{\uparrow}\bra{\uparrow}$.}
	\label{fig:appx_xxx_additional_numerical_results}
\end{figure}

\begin{figure}[ht]
	\centering
	\begin{minipage}{.45\textwidth}
		\centering
		\includegraphics[width=\columnwidth]{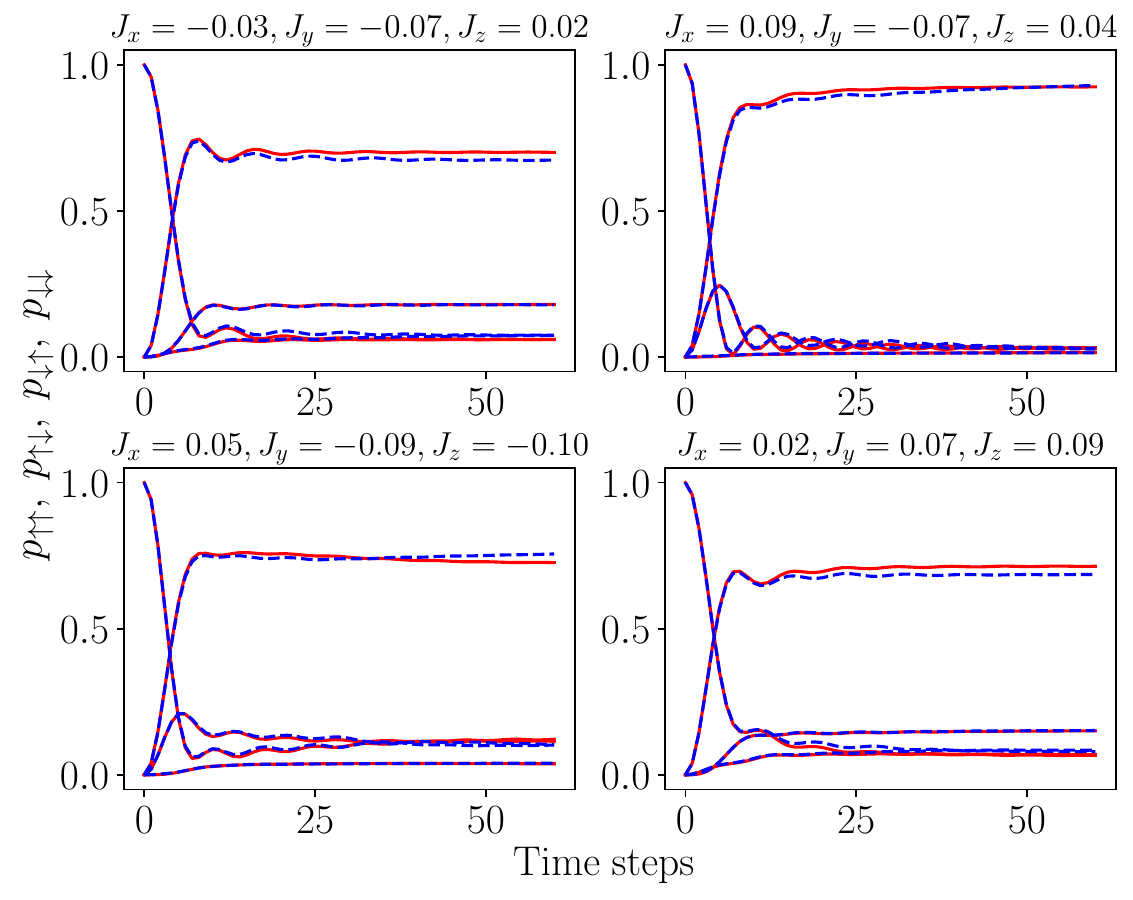}
	\end{minipage}
	\hspace{.05\columnwidth}
	\begin{minipage}{.45\textwidth}
		\centering
		\includegraphics[width=\columnwidth]{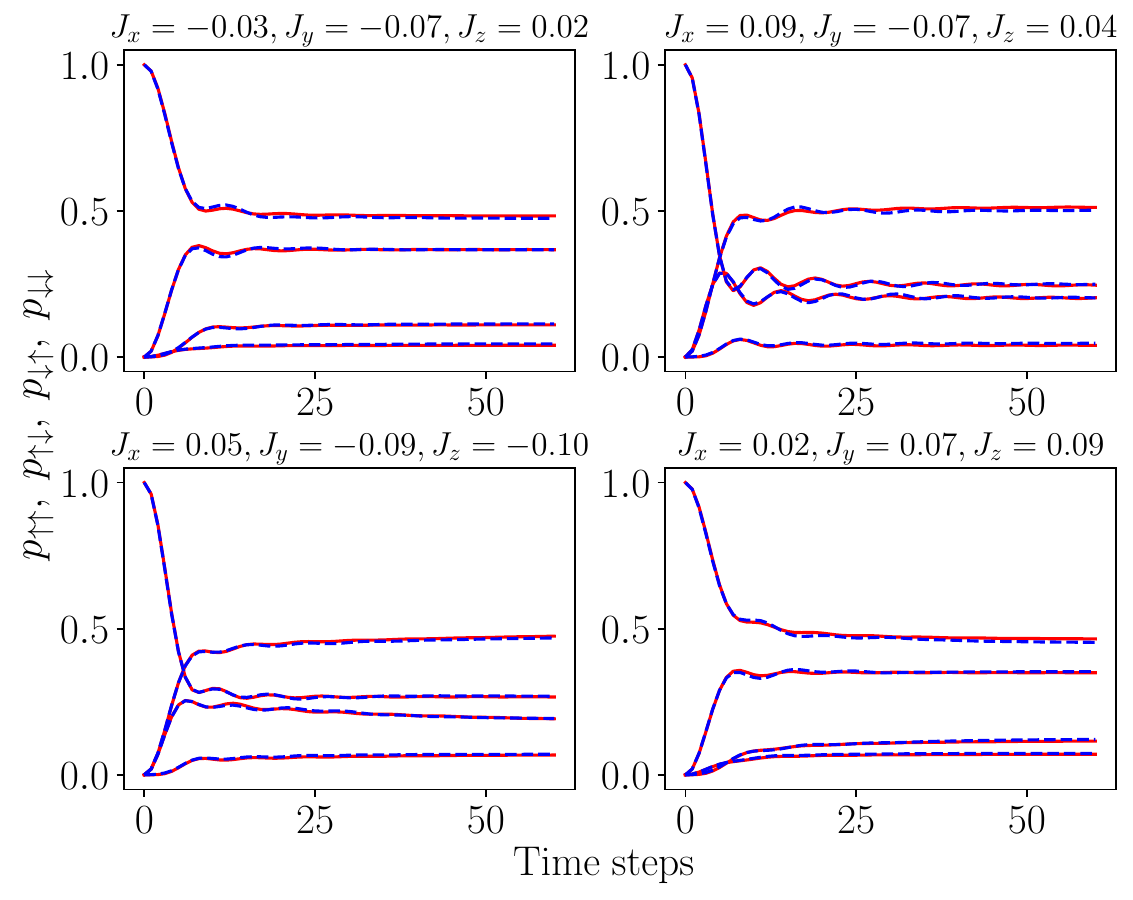} 
	\end{minipage}
	\caption{The comparison of the first-principles dynamics (solid lines) of the
	impurity density matrix diagonal elements, that are given by
	$p_{\uparrow\uparrow}=\bra{\uparrow}\otimes\bra{\uparrow}\rho^{\imp}\ket{\uparrow}\otimes\ket{\uparrow}$,
	$p_{\uparrow\downarrow}=\bra{\uparrow}\otimes\bra{\downarrow}\rho^{\imp}\ket{\uparrow}\otimes\ket{\downarrow}$,
	$p_{\downarrow\uparrow}=\bra{\downarrow}\otimes\bra{\uparrow}\rho^{\imp}\ket{\downarrow}\otimes\ket{\uparrow}$,
	$p_{\downarrow\downarrow}=\bra{\downarrow}\otimes\bra{\downarrow}\rho^{\imp}\ket{\downarrow}\otimes\ket{\downarrow}$,
	with the dynamics predicted using learned influence matrices (dashed
	lines). The left panel corresponds to XX environments with initial states being $\rho^L = \rho_{\downarrow}$ and $\rho^R = \rho_{\uparrow}$, the right panel corresponds to XX environments with initial states being $\rho^L = \rho_{\infty}$ and $\rho^R = \rho_{\uparrow}$. The size of datasets used to learn influence matrices is $N=7.5\cdot
	10^6$. The initial state of the impurity is $\rho^\imp(0) =
	\ket{\uparrow}\bra{\uparrow} \otimes \ket{\uparrow}\bra{\uparrow}$.}
	\label{fig:appx_xx_additional_numerical_resultss}
\end{figure}

In this appendix we give additional numerical results. Each additional
numerical experiment is classified by the type of left and right environments
(XX or XXX), initial states of left and right environment ($\rho^L$ and
${\rho^R}$), constants $J_x$, $J_y$ and $J_z$ defining a quantum channel acting on the impurity Eq.~\eqref{eq:impurity}. Constants
$J_x$, $J_y$ and $J_z$ are sampled uniformly from a cube $[-1, 1]^3$ and
rounded to two decimal places. For all the environments we set $J=0.1$. In
Fig.~\ref{fig:appx_xxx_additional_numerical_results} and
Fig.~\ref{fig:appx_xx_additional_numerical_resultss} we provide a comparison
of first-principles dynamics of diagonal elements of the impurity density matrix with the dynamics based on learned IMs.
\end{document}